\documentclass[modern]{aastex62}

\received{}
\revised{}
\accepted{}
\submitjournal{ApJ}

\shorttitle{FR0 radio galaxies and radio morphology}
\shortauthors{Garofalo \& Singh}

\begin{document}

\title{FR0 radio galaxies and their place in the radio morphology classification}

\correspondingauthor{David Garofalo}
\email{dgarofal@kennesaw.edu}

\author{David Garofalo}
\affil{Department of Physics, Kennesaw State University \\
Marietta GA 30060, USA}

\author{Chandra B. Singh}
\affiliation{The Raymond and Beverly Sackler School of Physics and Astronomy, Tel Aviv University,\\
Tel Aviv 69978, Israel}



\begin{abstract}

So-called FR0 radio galaxies have recently emerged as a family of active galaxies with all the
same properties as FRI radio galaxies except for their ratio of core to total emission which is
about 30 times as high compared to FRI sources. We show how their properties fit within the
gap paradigm as low prograde spinning black holes whose progenitors are powerful FRII quasars
that transitioned rapidly from cold mode into advection dominated accretion in a few million
years. The prediction is that if sufficient fuel exists, FR0 radio galaxies will evolve into full-
fledged FRI radio galaxies and the observational dearth of FRI radio galaxies compared to FR0s
at low redshift tells us about the supply of gas in the low redshift FR0s. Given the model
prescription, this 5 to 1 FR0 to FRI ratio implies that at low redshift the FRII quasar class of active
galaxies struggles to fuel its black hole beyond 1.3 times its original mass. In addition to this, we
illustrate model prescriptions for the black hole mass, black hole spin, redshift and environment
distribution for FR0 radio galaxies by fitting them within a paradigm that views them as a
continuous class of active galaxies that are sandwiched between FRII quasars and FRI radio
galaxies.

\end{abstract}



\section{Introduction} 

Although selection effects singling out active galactic nuclei (AGN) with powerful jets are responsible for our current classification scheme into FRI and FRII jets \citep{fan74},
deep large area surveys have revealed that compact, or small-scale jets, constitute the dominant population at least at low redshift \citep{bal09, bal10, sad14}.
These FR0 radio galaxies appear difficult to distinguish from the FRII and FRI low excitation radio galaxies (LERGs) \citep{cap17a, cap17b} but on closer inspection their line, X-ray luminosities and
core properties, show them to be environmentally indistinguishable from FRIs \citep{tor18, bal15}. Because the numbers of FRI radio galaxies are significantly smaller than
those of the FR0 radio galaxies, \citet{bal18a} conclude that FR0s cannot be thought of as young FRIs. In this work, however, we circumvent the criticism against
 FR0s evolving into FRIs by recognizing that FR0 radio galaxies need not be young objects, but accreting black holes in their middle ages, having evolved for hundreds of millions of years or more. 
As a result of their lack of youth, an evolutionary connection to extended jet emission as an FRI radio galaxy is plausible but not inevitable, as there must be sufficient amount 
of fuel to evolve the FR0 into its future FRI state. The observational discrepancy between the numbers of FRIs compared to those of FR0s is then interpreted as the result of a 
limit on the fuel of FR0s in the low redshift
universe. We will show that the giant black holes in the low redshift FRI radio galaxies fit into a paradigm in which they have been supplied with enough matter into their nuclei to build their
black holes to at least 30$\%$ more mass than when they were formed. In addition to this, we predict black hole mass and spins, as well as redshift and environment distribution for FR0s.\\

In next section we introduce the model. In Section 3 we illustrate how the model accommodates FR0s. In Section 4 we explore their environments. In Section 5 we connect them to their
immediate ancestor WLRG/FRIIs. In Section 6 we explore their black hole masses and redshift. And in Section 7 we conclude.

\section{The model}

The gap paradigm for black hole accretion and jet formation is a phenomenological scale-invariant framework for understanding black hole feeding and feedback that provides insight
into observations across the mass scale. The model has been applied to the most basic questions in AGN such as the radio loud/radio quiet dichotomy and the jet-disk connection by
opening a window on powerful jet production from retrograde accreting black holes \citep{gar10}; the redshift distribution of jetted AGN and compatibility with the Soltan
argument by recognizing a natural time evolution toward prograde spinning black holes \citep{gar16}; the nature of the Fundamental Plane \citep{gar14};
the redshift distribution of BL LACs and flat spectrum radio quasars \citep{garet18} and the environment dependence of radio galaxies as well as jet power/lifetime correlations
\citep{gar18}, among others. Despite a description in terms of simple cartoons to describe its time evolution, a surprising degree of quantitative predictability exists that in this
work will be applied to incorporate FR0 radio galaxies. \\ 
We outline the essential features of the model using Figures 1,2 and 3. Figure 1 is the result of the numerical solution of the force-free
equations for jet power as a function of black hole spin, with retrograde orientations of the disk captured by the minus sign and prograde orientations of the disk with positive signs.
With this figure we can identify regions of the spin space that correspond to relatively weaker jets, which will be identified with the FR0 radio galaxies. Figure 2 is new, introduced to
describe in greater time resolution the evolution of the most powerful FRII quasars but in a way that is directly connected to standard figures that have been used previously to apply the
model. Specifically, Figure 3 is less well resolved in time but constitutes a more general application of the model with a focus on the variety of different paths for FRII quasars. In
practical terms we should think of Figure 2 as included in Figure 3 as Figure 2 is a more detailed version of Figure 3c, the lowest series of panels in Figure 3, which describes the
predicted time evolution of the most massive and therefore most powerful FRII quasars that are born in gas rich mergers. For the small fraction of post-mergers that allow stable retrograde
accretion, the lowest panel in Figure 3c shows a powerful FRII jet emerging from a black hole accreting in cold mode, surrounded therefore by a thin, radiatively efficient, accretion disk. As a
direct consequence of the powerful jet feedback, the accretion disk transitions relatively rapidly into an advection dominated system (ADAF) on a timescale of about a few million years (see
applications of this evolution in \citet{gar18}). Hence, the second panel in Figure 2 shows an ADAF accreting black hole that has not yet been fully spun down so remains in
retrograde mode. This is equivalent to what is shown in Figure 3c. The continued accretion will spin such black holes down on timescales of hundreds of millions of years making them
transition toward a state characterized by low luminosity and a disappearing jet as the black hole spin approaches zero value. Such objects have been identified as the WLRG/FRIIs of
\citet{tad12}. For the purposes of this work, however, our interest is in the evolution toward a renewed jet state as the system evolves into the prograde spin regime. For fixed
magnetic field threading the inner disk $B_{d}$ , the jet power in the prograde regime increases as the spin increases from zero spin and tends to level off in the intermediate prograde regime (see
equation 1 and Figure 1). This comes about in the numerical solution as a result of the existence of the zero flux boundary condition in the gap region inside of the innermost stable
circular orbit (i.e. the Reynolds condition). As the spin increases in the prograde direction, the Blandford-Payne jet is weaker while the Blandford-Znajek jet is stronger. As a result they tend
to balance out and this produces a jet power vs spin that is less steep in the prograde regime compared to the retrograde one. At low prograde values, the jet is weaker but it increases
from zero spin and we identify a region of the jet power versus spin space that can model compact/weak FR0 radio galaxy jets. For simplicity we choose 0.1 as the value of the
dimensionless black hole spin to model FR0s. Near this relatively low value of spin the jet is not negligible yet not powerful enough to constitute an FRI LERG. Unlike Figure 3c, which shows
the state of the system at high prograde spin, our focus here is on the state of the system at low prograde spin, in a region of the parameter space where weak, compact jets, might be
prescribed. Hence, we focus on the a$\sim$0.1 region and produce a panel prescribing a weak jet emerging from the black hole (Figure 2, third panel from bottom). In short, Figure 2 is nothing
more than Figure 3c with an additional panel introduced between the retrograde panels and the high spin prograde panel.

\begin{equation}
L_{jet} = 5 \times 10^{47}ergs^{-1} f(a) (B_{d}/10^{5} G)^{2} m_{9}^{2},
\end{equation}

Equation (1) (from \citealt{gar10}) constitutes the jet power in terms of a function of spin f(a) capturing the effects of the gap region on the magnetic flux, the magnetic
field in the inner disk region, $B_{d}$, and the black hole mass in terms of 1 billion solar masses, $m_{9}$. In the next section we will further explore the implications of FR0s 
fitting in the paradigm as low spinning prograde accreting black holes.

\begin{figure}[ht!]
\plotone{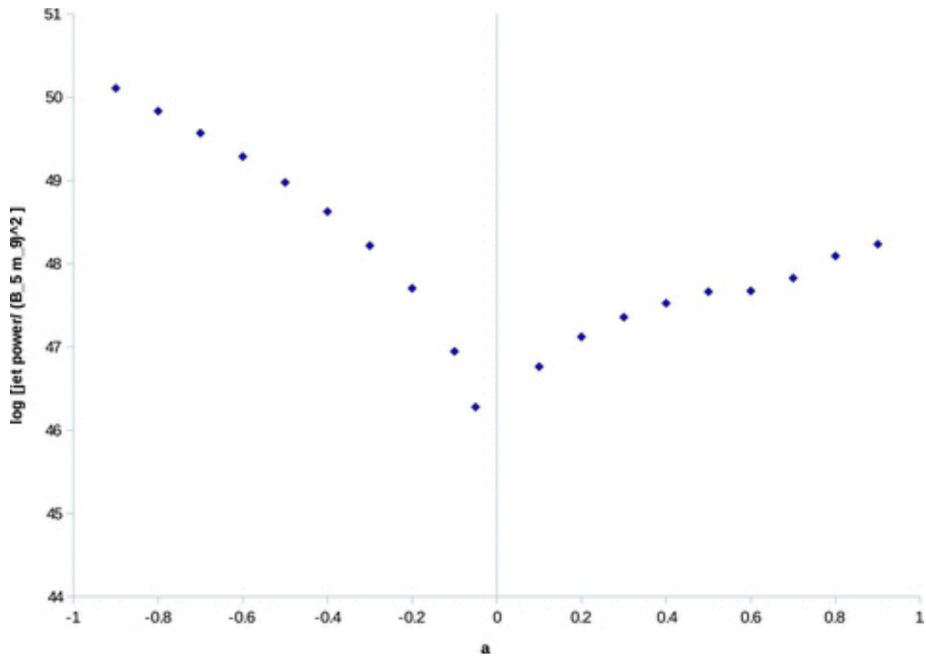}
\caption{Jet power as a function of the dimensionless black hole spin $a$ via equation 1 from (from \citealt{gar10}). At zero spin the jet must disappear and for near zero spin values 
the jet remains observationally undetected. For simplicity we choose 0.1 as the prograde value for which the jet is both visible but not yet a full-fledged FRI.}
\end{figure}

\begin{figure}[ht!]
\plotone{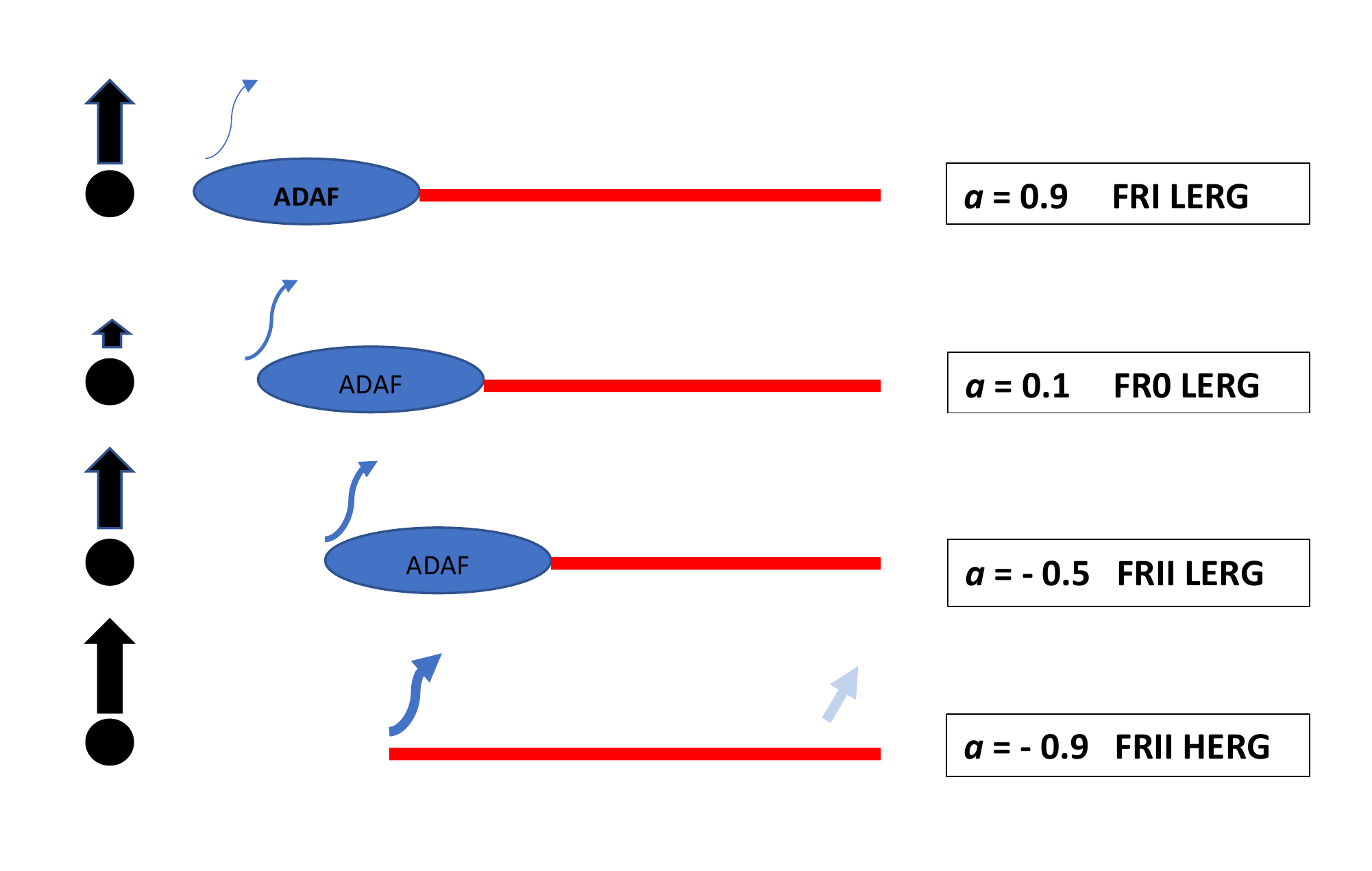}
\caption{Detailed time evolution of initially retrograde and high power FRII quasar (lowest panel). Accretion spins the black hole down toward zero spin but the 
accretion state evolves rapidly as a result of the powerful jet feedback. Hence the second to bottom panel shows a black hole that is still in retrograde configuration (a = -0.5)
but its disk has transitioned to an ADAF (blue vs red for thin disk). As the system transitions into the prograde
regime, the jet re-emerges but is weak due to the small value of the spin. This constitutes an FR0 radio galaxy in the
model. If sufficient fuel remains, the system becomes a full-fledged FRI LERG once the spin is sufficiently above 0.1
and the system remains an FRI LERG for as long as the black hole is fueled. The black arrows above the black hole
on the left represent the Blandford-Znajek jet with longer arrows representing more powerful jets, the curved
arrows at the disk inner edge represent the Blandford-Payne jet with thicker arrows representing more effective
jets (unlike in Figure 3 where the strength/effectiveness is captured by the length of the arrows), which are
progressively more effective as the spin increases in the retrograde direction, while the arrow in the disk represents
the disk wind, which is absent in ADAF states.}
\end{figure}

\section{FR0 radio galaxies}

As a result of the small value of the black hole spin, the jet is not powerful, and, like \citet{bal15} who identify the possibility that FR0s could be low 
black hole spin objects, we classify systems for which the spin is near the dimensionless value of 0.1 as belonging to the FR0 classification.
Continued accretion will eventually turn the FR0 LERG into an FRI LERG as can be seen in Figure 2. This can only occur, however, if sufficient fuel is available for the system to build its black hole
mass by the amount needed to spin it up to about 0.2 or beyond. A thin disk accumulates mass onto the black hole as determined by integrating the mass needed to shift the inner edge of the
disk $r$ in from the innermost stable circular orbit associated with the original high retrograde value to the final prograde value via equation (2) (\citealt{rai05, kim16}).

\begin{equation}
M_{acc} = \int dm/(1-2m/3r_{in})^{1/2}
\end{equation}

The expression is written in units such that Newton’s gravitational constant $G$ as well as the speed of light $c$ are equal to 1. $M_{acc}$ is the mass that is accreted onto the 
black hole, $m$ is the black hole mass, and $r_{in}$, as mentioned, represents the inner edge of the disk. If we follow the panels of Figure 2, this means $r$ in evolves from about 
8.7 gravitational radii (-0.9 spin) to about 5.3 gravitational radii (0.2 spin). Therefore, in order for the dying FRII LERG jet to re-emerge as an FRI LERG jet, the black hole 
must accumulate a mass of about 1.3 times its original mass. This can be seen in Figure 4 where the increase in black hole mass is shown as the black hole spins down and then 
up in the prograde regime. For the most massive systems of interest to us, the original black hole may be in the $10^{8}$ - $10^{9}$ solar mass range which means a minimum amount of
mass in the range of (0.03-0.3)$10^{9}$ solar masses must be accreted. The observational fact that at low redshift the numbers of FRI radio galaxies are only about $1/5$ of the total number of FR0
radio galaxies constitutes a constraint and prediction on the amount of material that is available to be accreted onto the black hole at low redshift in the environments of these low redshift FR0
radio galaxies. Unlike in \citep{bal18a}, however, it is important to recognize that we are indeed postulating an evolution from FR0s into FRIs, the difference that allows this to be
reasonable in our framework is that FR0s are not young objects but the product of prolonged accretion spanning timescales that easily accommodate hundreds of millions if not billions of
years. Given the amount of fuel that must be supplied to the black hole to accomplish the FR0 to FRI transition, however, we argue that it may be reasonable to imagine that not many low
redshift systems funnel sufficient fuel into their black hole sphere of influence to ultimately succeed. Hence, no expectation exists that all FR0 objects continue the project of generating
offspring FRI LERGs. Of course, these ideas are accessible to observational scrutiny.

\begin{figure}[ht!]
\plotone{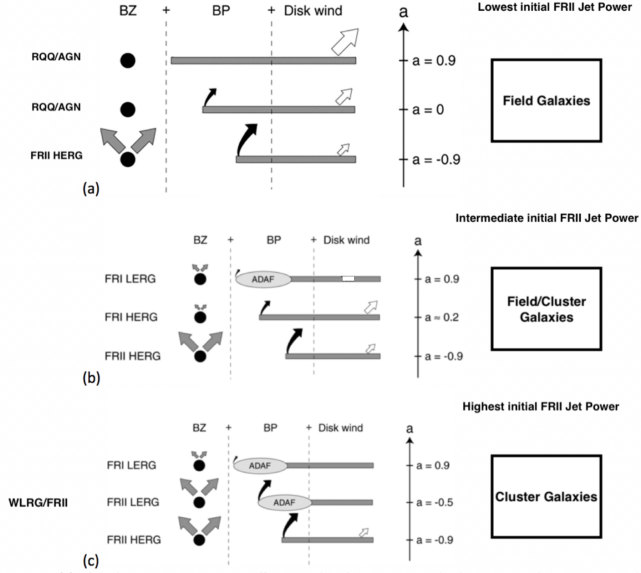}
\caption{Biggest (c) to smallest (a) black holes, and their environments (from \citealt{gar18}). The
FR0 radio galaxies emerge from the objects in (c) but for low, prograde spinning ADAFs which are not labeled here.
Such objects had FRII jets that died as the spin evolved progressively toward zero. These objects are prescribed to
evolve in rich clusters where the dark matter haloes are most massive.}
\end{figure}

\begin{figure}[ht!]
\plotone{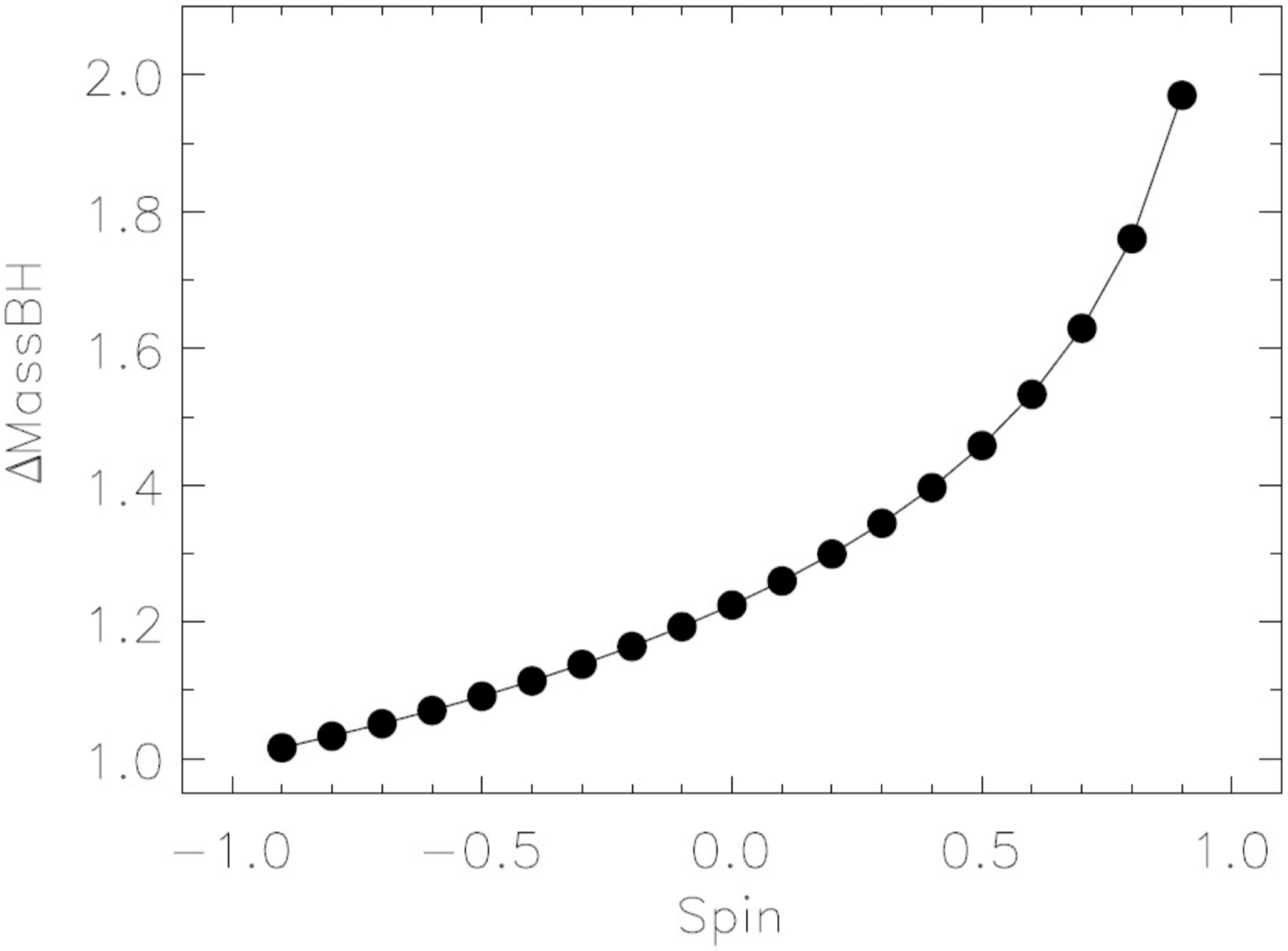}
\caption{Increase in relative mass for a black hole that begins accreting with high spin in the retrograde regime as
a function of spin (From \citealt{kim16}). If the black hole begins accreting at a retrograde spin of 0.9 and ends with
a prograde spin of 0.9, the black hole mass has roughly doubled.}
\end{figure}

\section{FR0 environments}

Recent work has shown how to understand the connection between the distribution of dark matter haloes in clusters versus field environments and the radio quasars and galaxies modeled
in the gap paradigm \citep{gar18}. In this section we apply those ideas to predict FR0s to be dominant in clusters as opposed to isolated fields. To appreciate this, we
invoke Figure 3 from \citet{gar18}, showing the time evolution and environments of radio quasars formed in mergers around black holes of different masses. In
Figure 3 (a) we have the smallest black holes whose jets are therefore relatively weaker, which makes them ineffective in altering the mode of accretion. Such black holes remain radiatively
efficient and therefore rapidly spin down and then up in the prograde direction. Because these accreting black holes are the smallest, they tend to be formed in isolated field galaxies whose
dark matter haloes are smaller. In Figure 3 (b) we see the time evolution of black holes with intermediate black hole masses. Such objects produce more effective jet feedback than in
Figure 3 (a) and therefore eventually experience a change in their accretion state from radiatively efficient to ADAF. These objects can be found both in field as well as in clusters. In
Figure 3(c), instead, we see the evolution of the most massive black holes. Such objects produce the most powerful and effective jet feedback heating their environment which constitutes the
kind of feedback that alters the accretion state rapidly into an advection dominated state. These are the progenitors to the FR0 radio galaxies. Since these are the most massive black holes, they
emerge in cluster environments where the most massive dark matter haloes live. The environment dependence of the three kinds of evolutionary paths described in Figure 3 have
been previously applied. The new element in this paper is the recognition that FR0s are an intermediate stage in the evolution of Figure 3c (i.e. Figure 2) and therefore are predicted to
emerge in environments where the largest dark matter haloes form, namely clusters.

\section{Progenitor WLRG/FRIIs}

By following the evolution of powerful FRII quasars or FRII HERGs in the paradigm via Figure 3 (c), we notice that the progenitors to the FR0 radio galaxies in terms of jets must be objects that
were low spinning retrograde black holes accreting in advection dominated form. These objects were modeled in \citet{gar18} as the WLRG/FRII systems explored in \citet{tad12}. 
FR0 radio galaxies are therefore the flip-side of the low retrograde spin objects and therefore are low spinning prograde black holes accreting in ADAF form. Because
there is an evolutionary link between these two subclasses of the radio galaxy population in the model, we expect the numbers of objects that fit within the WLRG/FRII classification, to be
more numerous compared to the FR0 radio galaxy group. Again, this comes about from the possibility that the amount of fuel available to be accreted onto the black hole may terminate
prior to reaching the low prograde spin values. Of course, on average, the redshift for the WLRG/FRII are larger than for the FR0 radio galaxies which produces different observational
challenges for the two subpopulations of radio AGN. In other words, the detection threshold introduces a bias in favor of the FR0s compared to the WLRG/FRIIs which appears compatible
with the observations in \citet{cap17b, bal18b} on FRIICAT and FR0CAT samples.

\section{Black hole mass and redshift}

The continuous accretion picture adopted here to model the FR0s requires that their black holes possess intermediate masses in the hierarchy of radio galaxies among the most massive black
holes. On average, i.e., among the most massive black holes, FRII quasars or FRII HERGs have the lowest black hole masses while the FR0 radio galaxies and FRI LERGs have increasingly larger
black hole masses on average. We have been precise about this in that continuous accretion provides the black hole with about an additional $30\%$ of its original mass for it to become a low,
prograde spinning, black hole, with sufficient spin to enter the FRI LERG classification. The model therefore prescribes FR0 LERG black hole masses to be on average no larger than about
$30\%$ more massive than the average for the FRII HERGs. For the fraction of FR0 LERGs that have enough fuel to spin their black holes up into the higher prograde regime to become FRI LERGs,
such objects will possess the largest black holes. Hence, in terms of their black hole masses, FR0 LERGs are sandwiched in-between FRII HERGs and FRI LERGs. This is in fact confirmed in the
studies on FRICAT and FR0CAT. In addition to a quantitative prescription for their black hole masses, the model allows us to also prescribe the redshift dependence of the various classes
of FRI galaxies described in this work. Restricting our focus on the most massive black hole accreting radio galaxies, the ones captured in Figure 3c and Figure 2, the transition to
ADAF/LERG states slows the evolution timescales down by at least two orders of magnitude compared to the objects described in Figure 3a. Hence, there is a non-negligible redshift gap
between the FRII HERGs of Figure 3c and Figure 2, and the FR0 LERGs of Figure 2, and an even greater redshift gap between the FRII HERGs and the high prograde spin FRI LERGs of Figure 3c
and Figure 2. For a recent application of the quantitative nature of that redshift gap between the FR radio galaxy subclass of AGN as applied to flat spectrum radio quasars and BL LACs, see 
\citet{garet18}.

\section{Conclusions}

This paper constitutes an application of the gap paradigm for black hole accretion and jet formation to the emerging class of FR0 radio galaxies. These objects 
live in environments thatare analogous to those of FRI radio galaxies in all respects but differ from them in their ratio of core to extended emission, which is 
tens of times larger. By fitting in the framework as low, prograde spinning black holes surrounded by advection dominated accretion flows, we have
argued that they are not young objects but the result of a renewed jet activity following the evolution of black hole spin whose past involved a retrograde spin down phase. As a result of
their advection dominated accretion disks, such objects struggle to spin their black holes up, and therefore require hundreds of millions of years if not order billions of years to reach the
spin values for which the paradigm prescribes non-negligible jets. Incidentally, this timescale is responsible for the appearance of powerful jets at sufficiently separated cosmic times 
\citep{garet18}. FR0 radio galaxies are thus imagined to be in their middle ages, such that if sufficient fuel allows them to be spun up further in the prograde direction, they will evolve into
more powerful jet producers, FRI radio galaxies, in their older age. We have shown that compatibility between the model and observations requires that most low redshift FR0 radio
galaxies do not have the ability to accrete beyond about $30\%$ of their original black hole mass, which may be explored observationally. We have also shown that because FR0 radio galaxies
were born as FRII quasars with the most massive black holes, this particular family of radio galaxy prefers rich cluster environments over isolated field galaxies, possessing intermediate
values of black hole mass among the most massive class and intermediate redshift values. Finally, we point out that FR0s do not fit in the paradigm as radio quiet quasars/AGN -
which the weakness or compactness of their jets might suggest - because their environments are not compatible with those for radio quiet quasars/AGN as seen in Figure 3(a) top two
panels.\\

\acknowledgements

We thank the anonymous referee for insights that improved the clarity and focus of the paper.

\end{document}